\documentclass[11pt]{article}\usepackage[]{graphicx}\usepackage[]{color}
\makeatletter
\def\maxwidth{ %
  \ifdim\Gin@nat@width>\linewidth
    \linewidth
  \else
    \Gin@nat@width
  \fi
}
\makeatother

\definecolor{fgcolor}{rgb}{0.345, 0.345, 0.345}

\usepackage{framed}
\makeatletter
 {\par\unskip\endMakeFramed%
 \at@end@of@kframe}
\makeatother

\definecolor{shadecolor}{rgb}{.97, .97, .97}
\definecolor{messagecolor}{rgb}{0, 0, 0}
\definecolor{warningcolor}{rgb}{1, 0, 1}
\definecolor{errorcolor}{rgb}{1, 0, 0}
\newenvironment{knitrout}{}{} 

\usepackage{alltt}

\usepackage{fullpage}
\usepackage{verbatim}
\usepackage{authblk}
\usepackage{amsmath}
\usepackage[left=1in,top=1in,right=1in,nohead]{geometry} 
\usepackage{setspace} 
\usepackage{url}
\usepackage[super,comma,sort&compress]{natbib} 
\usepackage[final]{changes}
\usepackage{graphicx}
\usepackage{grffile}
\usepackage{placeins} 

\IfFileExists{upquote.sty}{\usepackage{upquote}}{}
\begin{document}

\title{Real-time projections of epidemic transmission and estimation of vaccination impact during an Ebola virus disease outbreak in the Eastern region of the Democratic Republic of Congo}

\author[1]{Lee~Worden}
\author[1,2]{Rae~Wannier}
\author[3]{Nicole~A.~Hoff}
\author[3]{Kamy~Musene}
\author[4]{Bernice~Selo}
\author[4]{Mathias~Mossoko}
\author[5]{Emile~Okitolonda-Wemakoy}
\author[6]{Jean~Jacques~Muyembe-Tamfum}
\author[2]{George~W.~Rutherford}
\author[1,2]{Thomas~M.~Lietman}
\author[3]{Anne~W.~Rimoin}
\author[1,2]{Travis~C.~Porco}
\author[1,2]{J.~Daniel~Kelly\footnote{Corresponding author}}

\affil[1]{F.\ I.\ Proctor Foundation, University of California, San Francisco (UCSF), San~Francisco,~CA,~USA}
\affil[2]{School of Medicine, UCSF, San Francisco, CA, USA}
\affil[3]{School of Public Health, University of California, Los Angeles, Los Angeles, CA, USA}
\affil[4]{Ministry of Health, Kinshasa, Democratic Republic of Congo}
\affil[5]{School of Public Health, University of Kinshasa, Kinshasa, Democratic
Republic of Congo}
\affil[6]{National Institute of Biomedical Research, Kinshasa, Democratic Republic of Congo}

\maketitle

\doublespace

\maketitle

\newpage

\begin{abstract}
\noindent
\textbf{Background:} As of October 12, 2018, 211 cases of Ebola virus disease (EVD) were
reported in North Kivu Province, Democratic Republic of Congo. 
Since the beginning of October the outbreak has largely shifted into
regions in which active armed conflict is occurring, and in which EVD
cases and their contacts are difficult for health workers to reach.
We used available data on the current outbreak with
case-count time series from prior outbreaks to project the short-term
and long-term course of the outbreak.
\\
\textbf{Methods:}
For short and long term projections
we modeled Ebola virus transmission using a stochastic branching
process that assumes gradually quenching transmission estimated from
past EVD outbreaks, with outbreak trajectories conditioned on
agreement with the course of the current outbreak,
and with multiple levels of vaccination coverage.
We used a negative binomial autoregression for short-term projections,
a Theil-Sen regression model for final sizes,
and a baseline minimum-information projection using Gott's law
to construct an ensemble of forecasts to be compared and
recorded for future evaluation against final outcomes.
From August 20 to October 13, short-term model projections
were validated against actual case counts.
\\
\textbf{Results:}
During validation of short-term projections, from one week to four weeks,
we found models consistently scored higher on shorter-term forecasts.
Based on case counts as of October 13,
the stochastic model projected a median case count of
226 cases by October 27
(95\% prediction interval: 205--268)
and
245 cases by November 10
(95\% prediction interval: 208--315),
while the auto-regression model projects median case counts of
240
(95\% prediction interval: 215--307)
and
259
(95\% prediction interval: 216--395)
cases for those dates, respectively.
Projected median final counts range from 274 to 421.
Except for Gott's law, 
the projected probability of an outbreak comparable to
2013--2016 is exceedingly small.
The stochastic model estimates that vaccine coverage in this outbreak
is lower than reported in its trial setting in Sierra Leone.
\\
\textbf{Conclusions:}
Based on our projections we believe that the epidemic had not yet peaked at the
time of these estimates, though a trajectory on the scale of the West African
outbreak is exceedingly improbable. Validating our models in real time allowed
us to generate more accurate short-term forecasts, and this process may provide
a useful roadmap for real-time short-term forecasting. We estimate that
transmission rates are higher than would be seen under target levels of 62\%
coverage due to contact tracing and vaccination, and this model estimate may
offer a surrogate indicator for the outbreak response challenges.
\end{abstract}

\section*{Introduction}

On August 1, 2018, the World Health Organization (WHO) announced a new outbreak
of Ebola virus disease (EVD) in North Kivu Province in
Eastern Democratic Republic of Congo (DRC).
Epidemiological investigations traced EVD cases back to the week of April 30
and identified the initial epicenter to be Mabalako. North Kivu has over eight
million inhabitants, some of whom suffer from armed conflict, humanitarian
crisis, and displacement from their bordering countries of Uganda and Rwanda.
Since the outbreak began, EVD cases have spread across ten health zones in two
provinces at a rate outpacing the case counts of the other 2018 DRC Ebola
outbreak in Equateur Province.
As of October 13, 211 EVD cases were reported (31 probable and 180
confirmed); Ministry of Health of DRC, World Health Organization, and other
organizations were responding to the Ebola outbreak. 

As new interventions such as vaccines or rapid diagnostics are being
implemented during outbreaks, their impact on epidemic transmission is poorly
understood, requiring assumptions to be made that may lead to inaccurate
forecasting results.
Unknown social or environmental differences affecting transmission can
also affect forecasts in unknown ways.
For example, the overlap of the outbreak with regions where armed conflict
is occurring in North Kivu, DRC, might result in higher
under-reporting rates and lower vaccine coverage than in other outbreaks,
causing increased transmission and decreased accuracy of reporting,
or might result in reduced transmission due to reduced mobility or other
considerations.
Since the beginning of October, an increased rate of detection of
new cases has been observed in the conflict zone,
perhaps due to reduced disease control.


During an Ebola outbreak, real-time forecasting has the potential to support
decision-making and allocation of resources, but highly accurate forecasts
have proven difficult for Ebola \cite{RW7,RW2} as well as other diseases.%
\cite{RW3,RW5,RW17,RW1}
Moreover, there are mathematical reasons to believe that
highly accurate forecasts of small, noisy outbreaks may never be
possible.\cite{RW11}
Nevertheless, while predicting the exact number of cases is unlikely
to ever be possible, forecasts which are accurate enough to be useful
may be possible.
Previous work has found that probabilistic forecasts can have
relatively high accuracy within a few weeks, though they become
less useful as time horizons grow longer \cite{Funk177451},
and short-term forecasts may provide useful information
for response organizations.


In this paper we apply a suite of independent methods of real-time
forecasting to the Eastern DRC outbreak, to generate both
short-term and long-term projections of future case counts
as of the time of writing.
We validate short-term projections by scoring projections
derived from case count reports obtained earlier in the outbreak
against subsequent known counts.
We include past and present projections (in supplemental material)
for future evaluation.
We summarize model results in terms of projections of the future
course of the outbreak, and interpret their implications
relevant to current rates of transmission and vaccine coverage
in conflict zones and overall.

\section*{Methods}

We used four techniques to derive real-time projections of
future case counts:
a stochastic simulation model calibrated to time-dependent
transmission rates measured from past outbreaks of EVD and
constrained to the observed partial trajectory of the current outbeak,
extending the model used in our previous work on the Spring outbreak;
a negative binomial auto-regression model predicting the course of
the outbreak from its course to date together with the course
of previous outbreaks;
a regression model for final size based on past outbreaks;
and a simple final size projection using Gott's law, which assumes only that
the proportion of the outbreak observed so far is entirely unknown.

\subsection*{Data sources}

Data on the current outbreak was collected from the WHO website in real time
as updated information was published.\cite{RW23}
A cumulative case count of probable and confirmed cases was generated
to be consistent with the best knowledge at the time.
Copies of the list of case counts were kept as of multiple dates
(Figure~\ref{fig:case-counts}), to be used in retrospective
scoring of model projections against subsequently known counts.
Though the epidemic was officially reported in late July
as a cluster of cases occurring in June and July,
seven sporadic early cases from April and May
were later linked to the current outbreak and added to later case totals.
This additional knowledge was added retrospectively
to the time series of cumulative case counts
only for predictions made for days on or after September 15th,
when these cases were officially linked to the current outbreak.  




\subsection*{Stochastic model}

We modeled Ebola virus transmission using a stochastic branching process model,
parameterized by transmission rates estimated from the dynamics of prior EVD
outbreaks, and conditioned on agreement with reported case counts from the 2018
EVD outbreak to date. We incorporated high and low estimates of vaccination
coverage into this model. 
We used this model to generate a set of probabilistic projections
of the size and duration of simulated outbreaks in the current setting. 
This model is similar to one described in previous work \cite{Kelly331447},
with the addition of a smoothing step 
allowing transmission
rates intermediate between those estimated from previous outbreaks.

To estimate the reproduction number $R$ as a function of the number of days from
the beginning of the outbreak, we included reported cases by date from fourteen
prior outbreaks (Table~\ref{table:past-outbreaks}).%
\cite{RN711,RN371,RN712,RN695,RN714,CDC2001,RN715,BF2005,WHO2005,RN718,RN710,RN716,WHO2012,CDC2017}
The first historical outbreak reported in each
country was excluded (e.g., 1976 outbreak in Yambuko, DRC).
As there is a
difference in the Ebola response system as well as community sensitization to
EVD following a country’s first outbreak,
we employed this inclusion criterion
to reflect the Ebola response system in DRC during what is now its tenth
outbreak.
We used the Wallinga-Teunis technique to estimate $R$ for each case and
therefore for each reporting date in these outbreaks.\cite{RN727}
The serial interval
distribution used for this estimation was a gamma distribution with a mean of
14.5 days and a standard deviation of 5 days, with intervals rounded to the
nearest whole number of days, consistent with the understanding that the serial
interval of EVD cases ranges from 3 to 36 days with mean 14 to 15 days.
We estimated an initial reproduction number $R_\text{initial}$
and quenching rate $\tau$
for each outbreak by fitting
an exponentially quenched curve
to the outbreak's estimates of $R$ by day $d$
(Figure~\ref{fig:R-tau-fits}).

We modeled transmission using a stochastic branching process model in which the
number of secondary cases caused by any given primary case is drawn from a
negative binomial distribution whose mean is the reproduction number $R$
as a function of day of the outbreak, and
variance is controlled by a dispersion parameter $k$.%
\cite{RN705,RN706}
All transmission events were assumed to be independent.
The interval between date of detection of each
primary case and that of each of its secondary cases is assumed gamma
distributed with mean 14.5 days and standard deviation 5 days, rounded to the
nearest whole number of days, as above.

We used the $(R_\text{initial},\tau)$ pairs estimated from past outbreaks
to provide $R$ values for simulation.
$R_\text{initial}$ values were sampled uniformly from the range of values
estimated from past outbreaks.
We fit a linear regression line through the values of $R_\text{initial}$ and
$\log(\tau)$ estimated for past outbreaks, above,
and used the resulting regression line to assign a mean $\tau$
to each $R$, used
with the residual variance of $\log(\tau)$ as a distribution from which
to sample $\tau$ values for simulation given $R_\text{initial}$.
The pair of
parameters $R_\text{initial}$ and $\tau$ sampled in this way, together with
each of three values of the
dispersion parameter $k$, $0.3$, $0.5$, and $0.7$, consistent with
transmission heterogeneity observed in past Ebola outbreaks,
were used to generate simulated outbreaks.

This model generated randomly varying simulated outbreaks with a range of case
counts per day. The outbreak was assumed to begin with a single
case. The simulation was run multiple times, each instance producing
a proposed epidemic
trajectory, generated by the above branching process with
the given parameters $R_\text{initial}$, $\tau$, and $k$, and
these were then filtered by discarding all proposed outcomes but those whose
cumulative case counts matched known counts of the current 2018 EVD
outbreak on known dates.
In earlier, smaller, data sets we filtered against all reported
case counts,
while in later, more complete data sets we thinned the case counts,
for computational tractability,
by selecting five case counts evenly spaced in the data set plus
the final case count (Figure~\ref{fig:case-counts}).
The filtration required an exact match of the first target value,
and at subsequent target dates accepted epidemics within a number of
cases more or less than each recorded value.
On the earlier data sets in which the beginning dates of the epidemic
were unknown, the first target value was allowed to match on any day,
and subsequent target dates were assigned relative to that day.

Thus this model embodies a set of assumptions that
transmission rates are overall gradually declining from the start of the
outbreak to its end, though possibly in noisy ways.
When the tolerance of the filter on case counts is small,
quenching of transmission through time must closely track
case counts, while when tolerance is high, fluctuations in
the rate of generation of new cases can reflect a pattern of
ongoing quenching of transmission more loosely and on the long
term, while being more insensitive to
short-term up and down fluctuations in transmission rates
reflected by the true case counts.

We varied the tolerance as the data set became more complete to
maintain a roughly fixed rate of generation of filtered trajectories:
on the August 20 data set we allowed a tolerance of 4 cases more or less
than each target count,
on August 27 and September 5, 6 cases,
on September 15, 10 cases,
on October 7, 12 cases,
and on October 13, 17 cases.
This one-step particle filtering
technique produced an ensemble of model outbreaks, filtered on agreement with
the recorded trajectory of the outbreak to date.
This filtered ensemble was then
used to generate projections of the eventual outcome of the outbreak.\cite{RW13}

To model vaccination coverage with respect to total transmission (unreported
and reported), we multiplied the estimate of vaccine effectiveness by low and
high estimates of reported cases. In a ring vaccination study at the end of the
West Africa outbreak, the overall estimated rVSV-vectored vaccine efficacy was
100\% and vaccine effectiveness was 64.6\% in protecting all contacts and
contacts of contacts from EVD in the randomized clusters, including
unvaccinated cluster members.\cite{RN734}
We used estimates of vaccine effectiveness in our
stochastic model. The ring vaccination study found the vaccine to be effective
against cases with onset dates 10 days or more from the date of vaccine
administration, so we modeled the vaccination program as a proportionate
reduction in the number of new cases with onsets 10 days or more after the
program start date. 

We used past estimates of the proportion of unreported cases to estimate
the proportion of exposed individuals not covered by the vaccination process.
Based on a Sierra Leonean study from the 2013--2016 outbreak,\cite{RN733}
we estimated that
the proportion of reported cases in DRC would rise over time from a low of 68\%
to a high of 96\%. Given these low and high estimates of reported cases and the
estimate of vaccine effectiveness, a low estimate of vaccination program
coverage was 44\% ($68\% \times 64.6\%$) and a high estimate of vaccination program
coverage was 62\% ($96\% \times 64.6\%$).
We modeled the course of the outbreak with
and without the vaccination program based on approximate dates available from
situation reports.\cite{RW23}

For simulation based on cases as of October 13, 320 outbreaks
were retained from 34,663,104 simulated outbreaks
after filtering on approximate agreement with DRC case counts. 
(Numbers of simulations from earlier data sets are reported in
Supplemental Materials.)
The simulated outbreaks that were retained after filtering
were continued until they generated no further cases.
Rare simulated outbreaks that exceeded 300,000 cases were capped at the
first value reached above that number, to avoid wasted computation.
We used this ensemble to
derive a distribution of final outbreak sizes, and of cumulative counts
at specific forecasting dates.
Projection distributions were derived using kernel density estimation
with leave-one-out cross-validation to determine bandwidth,
using a log-normal kernel for final sizes, due to the extended
tail of the values, and a normal kernel for all other estimates.
We calculated 
median values and 95\% prediction intervals using the 2.5 and 97.5 percentiles
of simulated outbreak size and duration.
We conducted the analyses using R
3.4.2 (R Foundation for Statistical Computing, Vienna, Austria). 

\subsection*{Auto-Regression model}

A negative binomial autoregressive model was chosen through a validation
process to forecast additional new case counts at time points one week, two
week, four weeks, and two months from the current date.  To adjust for
disparities in the frequency of case reporting in historic outbreaks, the data
were weighted by the inverse square root of the number of observations
contributed to the model. Models considered included parameters for historic
raw case counts at different time points, logs of raw case counts, ratio of
historic case counts to try and capture the trend of the epidemic curve,
log(time), and an offset for current case total.  When historic case counts for
specific dates were missing, each missing case count was linearly interpolated
from the two nearest case counts, allowing the model to remain agnostic about
the current trend of the epidemic.  After model fitting and validation, the
final model chosen was a log-link regression for additional cases on the number
of new cases identified in the previous two weeks, the previous four weeks and
the ratio of these two case counts.  

\subsection*{Regression model}

We conducted a simple regression forecast based solely on outbreaks of
size 10
or greater, based on prior outbreaks.\cite{RN711,RN371,RN712,RN695,RN714,CDC2001,RN715,BF2005,WHO2005,RN718,RN710,RN716,WHO2012,CDC2017}
Nonparametric Theil-Sen regression (R package \texttt{mblm}) was used to
project the final outbreak size
based on values of the outbreak size at a specific earlier time. All time series
were aligned on the day they reached 10 cases.  
Finally, we reported the median and 95\% central coverage intervals for the
prediction distribution, conditional on the predicted value being no smaller than the
observed value for each day. Full details are given in \cite{Kelly331447}.
All analyses were conducted using R 3.4.2
(R Foundation for Statistical Computing, Vienna, Austria).

\subsection*{Gott's law model}

With
Gott's Law, we assume we have no special knowledge of our position on the
epidemic curve.\cite{gott1993implications} 
If we assume a non-informative uniform prior for the portion $\alpha$
of the epidemic included in the last available report,
the corresponding probability density function for the final size
$\mathrm{Y}=Y_0/\alpha$
is $Y_0/y^2$, $Y_0\leq y$.
We constructed a probability mass function by assigning all probability
density to the whole number of days given by the integer part of each value.
We used this probability mass function as a projection of the final outbreak
size.

\subsection*{Scoring}

Each of the above models was used to generate an assignment of probability
to possible values of multiple quantities:
\begin{itemize}
\item Case count 1 week after the last available case count
\item Case count 2 week after the last available case count
\item Case count 4 weeks after the last available case count
\item Case count 2 months after the last available case count
\item Final outbreak size
\end{itemize}

Each model's performance on each of these projections was scored by
recording the natural logarithm of the probability it assigned to
the subsequently known true value of the quantity in question.

The short-term projections based on real-time
reporting were used to evaluate and calibrate the models during
the epidemic, based on the data available
at multiple time points during the outbreak.
Final outbreak size projections were recorded
for future evaluation of their performance.

\section*{Results}

\begin{figure}
\centering
\begin{tabular}{ll}
Data as of 8-20-2018 & Data as of 8-27-2018 \\
\includegraphics[width=0.4\textwidth]{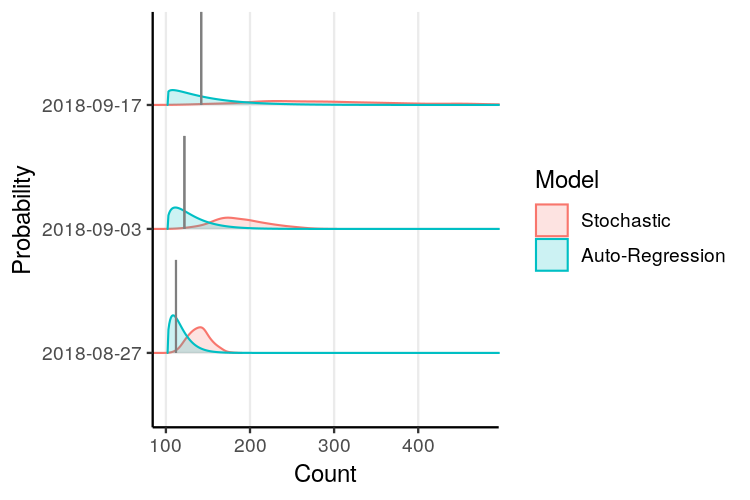} &
\includegraphics[width=0.4\textwidth]{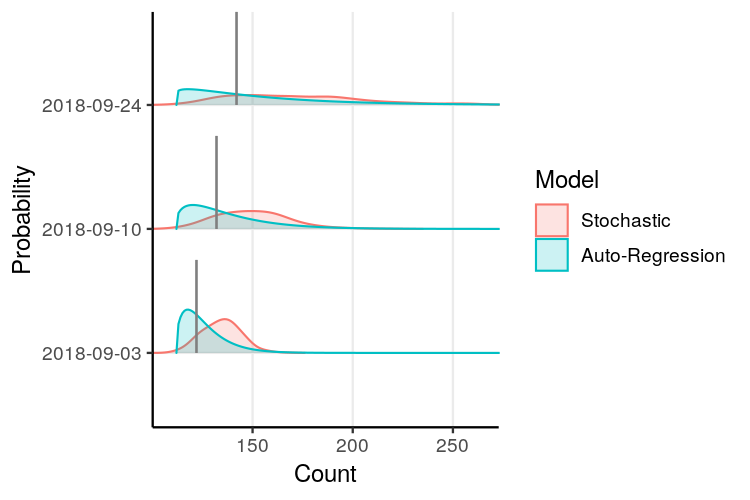} \\
Data as of 9-5-2018 & Data as of 9-15-2018 \\
\includegraphics[width=0.4\textwidth]{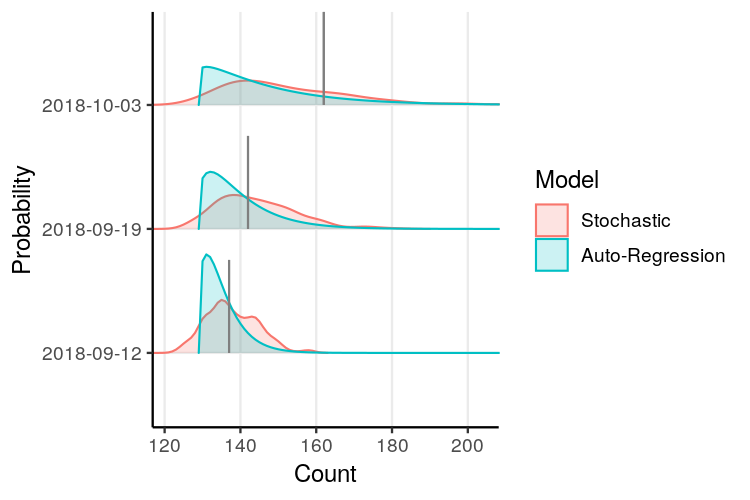} &
\includegraphics[width=0.4\textwidth]{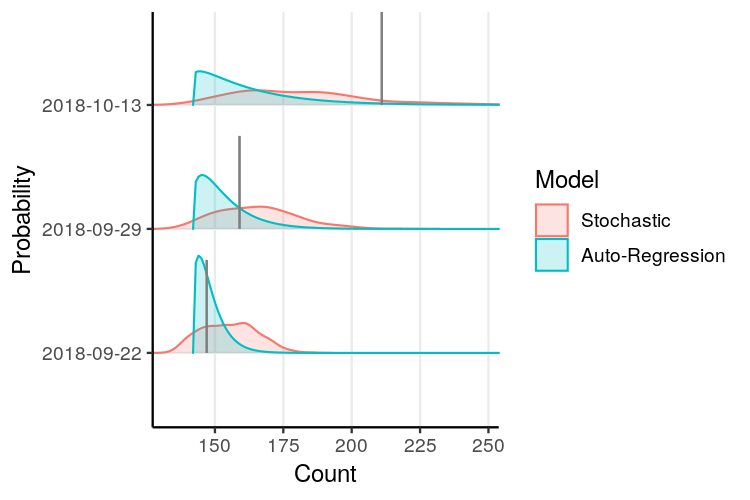} \\
\end{tabular}
\caption{\label{fig:scored-projections}
\textbf{Comparison of retrospective model projections
to known case counts}
when projecting from past snapshots of available data.
}
\end{figure}

\begin{figure}
\centering
\includegraphics[width=0.7\textwidth]{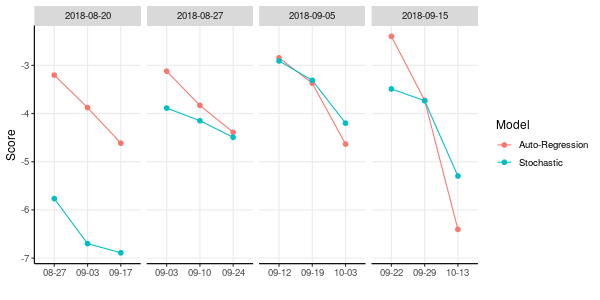}
\caption{\label{fig:scores}
\textbf{Log-likelihood scores of retrospective model projections on known case counts.}
}
\end{figure}

\begin{figure}
\centering
\includegraphics[width=0.5\textwidth]{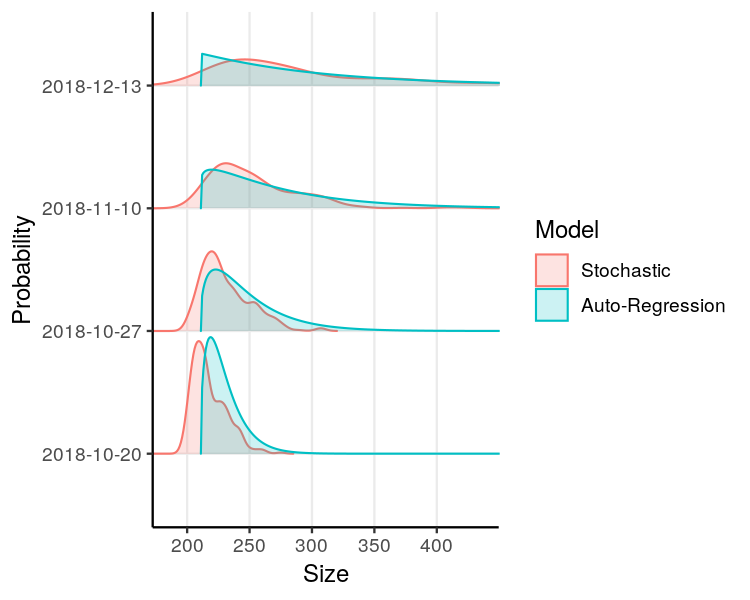}
\caption{\label{fig:short-term-forecasts}
\textbf{Short-term projections of case counts}
based on reported counts as of Oct. 13.
}
\end{figure}

\begin{figure}
\centering
\includegraphics[width=0.5\textwidth]{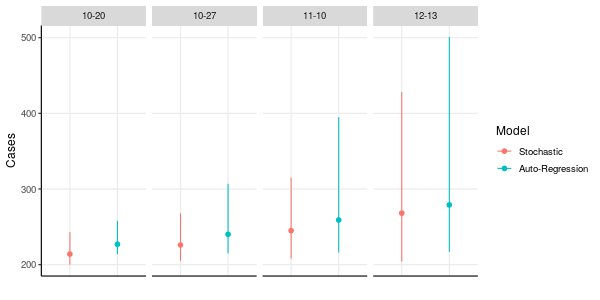}
\caption{\label{fig:short-term-boxplots}
\textbf{Medians and prediction intervals from
short-term projections of case counts}
based on reported counts as of Oct. 13.
}
\end{figure}

\begin{figure}
\centering
\includegraphics[width=0.5\textwidth]{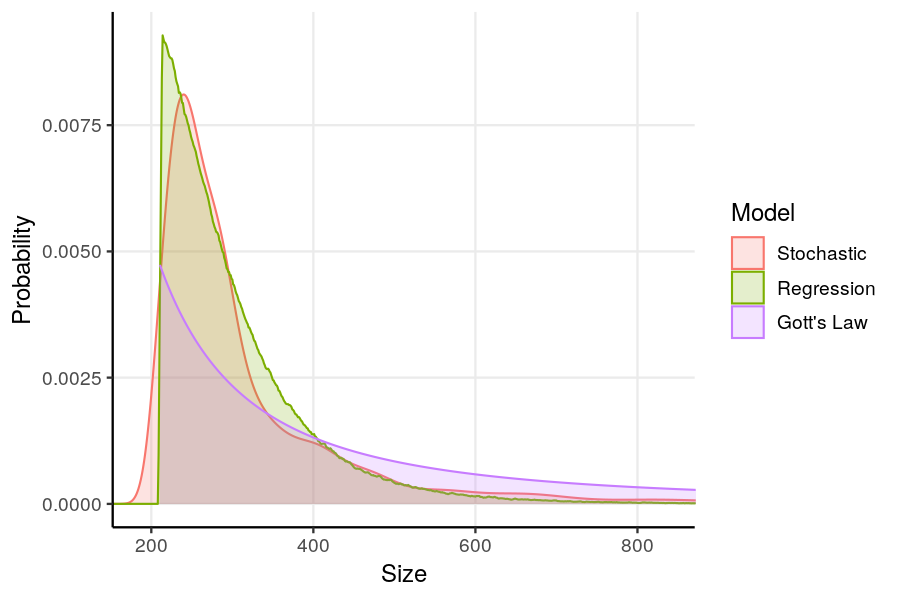}
\caption{\label{fig:final-forecasts}
\textbf{Projections of final case counts}
based on reported counts as of Oct. 13.
}
\end{figure}

\begin{figure}
\centering
\includegraphics[width=0.5\textwidth]{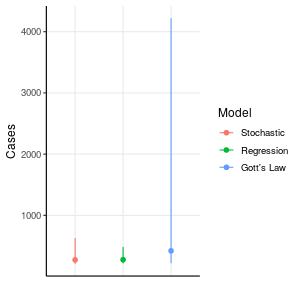}
\caption{\label{fig:final-boxplots}
\textbf{Medians and prediction intervals from projections of final case counts}
based on reported counts as of Oct. 13.
}
\end{figure}

When we started performing our short-term forecasts on August 20, 2018,
there were 102 reported EVD cases in North Kivu and Ituri provinces, DRC.
We used our stochastic and auto-regression models to project
one-week, two-week, four-week, and two-month forecasts of outbreak size.
As time lapsed, we compared predicted and actual outbreak sizes
and found a higher probability of accurate forecasts
at one week than two months
(Figures~\ref{fig:scored-projections},~\ref{fig:scores}).
Log-likelihood scores typically declined
as the model extended its projection time into the future.
However, the largest decline in log-likelihood score occurred
between the four-week and two-month forecasts.
Concurrently, there were larger prediction intervals
associated with these longer-term forecasts.
As the epidemic curve accelerated in early October,
we observed that model projections were less likely
to predict actual case counts.
These findings were consistent for both models.

After our model validation process was completed,
we used the stochastic and auto-regression models to project
one-week, two-week, four-week, and two-month outcomes
(Figures~\ref{fig:short-term-forecasts},~\ref{fig:short-term-boxplots}).
We used the Gott's law and Theil-Sen regression models
together with the stochastic model
to project final outbreak sizes
(Figures~\ref{fig:final-forecasts},~\ref{fig:final-boxplots}).
As of October 13, there were 216 reported EVD cases.
With the stochastic model,
the four-week projection of median outbreak size was
245 cases
(95\% prediction interval:
208--315).
Median final outbreak size was
274 cases
(95\% prediction interval: 
210--632).
With the auto-regression model,
the four-week projection of median outbreak size was
259 cases
(95\% prediction interval: 216--395).
With Gott's law, median final outbreak size was
421 cases
(95\% prediction interval: 222--4219).
Median final outbreak size projected by the regression model was
277 cases
(95\% prediction interval: 216--485).

Because the question has been raised of whether the current outbreak
might exceed the catastrophic West Africa outbreak in size,
we evaluated the model's projected probability of a final size
of at least the 28,616 cases reported in that outbreak.\cite{CDC2017}
A final outbreak size of 28,616 or more cases was projected to have
an exceedingly low probability of less than 1 in 10,000 in all
cases except the Gott's law model, whose projected probability distribution
is very long-tailed, which projects a probability of about 0.005
(roughly 1 in 190) for that event.
However, as with all of the above projections, it should be understood
that they are conditional on model assumptions being met.
If unpredictable events should change patterns of transmission,
for example escape of the outbreak into a region where sustained
high transmission rates violate the assumption of gradual
quenching of transmission, model projections will no longer
be applicable.

\subsection*{Stochastic model}

In order to produce model outbreak trajectories consistent with
the reported overall case counts since the beginning of October,
it was necessary to make the filtering step of the model more
tolerant to variation in counts in order to accommodate the
rapidly rising count.
This is because higher transmission rates in late September and
early October were necessary to generate case counts of that size
than are consistent with the earlier counts.

If this model's assumptions of continually quenching overall
rate of transmission is accepted, this result could be taken as
evidence in favor of increased transmission in the conflict
zone, since the increase in cases reported in October reflects
cases recorded there.

\begin{figure}
\centering
\includegraphics[height=0.4\textwidth]{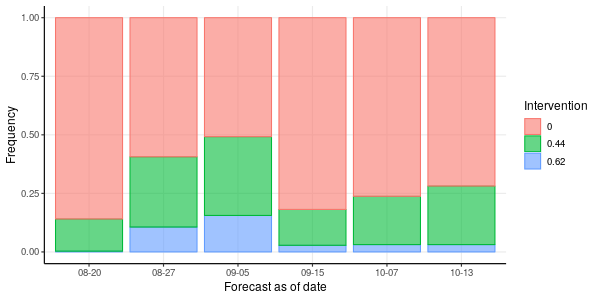}
\caption{\label{fig:vaccine-likelihoods}
\textbf{Likelihoods of vaccine coverage scenarios}
estimated by number of simulated outbreaks accepted by
the stochastic model's filtering step in which simulated outbreaks
must match reported case counts.
}
\end{figure}

The likelihoods of the
three scenarios of zero, low, and high vaccine coverage
estimated by the stochastic model,
on the basis of which scenarios are selected by the filtering step,
indicate that the lower vaccine coverage scenario
was consistently found more likely than the higher vaccine coverage scenario
(Figure~\ref{fig:vaccine-likelihoods}).
However, no vaccine coverage was the most likely scenario in all
forecasts.

This could be read as evidence for decreased vaccine coverage in the
conflict zone. It should be said clearly, however, that the model's
quenching assumption could be violated by the presence of other causes
of increased transmission relative to past outbreaks
that it can not distinguish and would
misidentify as decreased vaccine coverage.

Stochastic model parameters conditioned on filtering by true case counts,
stochastic model outcomes for multiple snapshots of reported data,
and short-term and long-term projections generated by all models
from past snapshots are reported in Supplemental Material.

\FloatBarrier

\section*{Discussion}

Political instability, mobility and community impenetrability to
health workers in
Eastern DRC present new challenges to efforts to respond to the ongoing EVD
outbreak. Public health responders have not been able to trace up to 80\% of
contacts of EVD cases,
and new chains of transmission are being identified on a
routine basis. At present, the most reliable source of data is the weekly case
counts that can be found in the WHO situation reports; those have indicated
that the number of additional cases is increasing rather than decreasing.
In such situations of data scarcity, modeling in real time can be useful,
particularly in the short term.

Our
projection with multiple models, as of October 13, 2018, predicts growth in
the short term consistent with rates recorded in the recent past.
In the long term our models do not project 
large-scale growth of the outbreak into a public health
emergency of international concern, even considering worst case scenarios.
However, these outcomes are likely
dependent on highly contingent events such as escape of the
outbreak into additional regions or nations which can not be predicted
by mechanistic models.

This is the first EVD outbreak in a conflict zone,
and the first with deployment of the vaccine since the beginning
of the outbreak,
so the suitability
of prior mathematical models to predict the course of this
particular outbreak is unknown.
We validated the short-term forecasting
performance of two mathematical models during the early part of the ongoing
outbreak.

Our short-term projections of the future course of the outbreak
at the time of writing are in rough agreement with expert
consensus that the outbreak has gained speed as cases appear
in the conflict zone, and will likely continue at roughly the
same rate of growth in the short term.
Our models do not indicate that the epidemic has yet peaked.
Longer term forecasts are less certain and depend on whether
intervention, conflict and mass behavior are able to
stem continued transmission.

There are limitations to our projections. Projection distributions are
right-skewed, with long tails (and we therefore report the median instead of
the mean).
We were unable to include all the 23 observed EVD outbreaks
with a case count greater than ten cases in our estimates due
to data availability.
The simple regression projection is
based entirely on past outbreaks of Ebola virus disease (measured and reported
in different ways), and cannot account for the improved control measures and
vaccination in the way that a mechanistic model does.
We included as
much real-time information into our estimates as possible, but situations such
as the introduction of EVD into a zone of armed conflict and the recent
introduction of vaccination are not reflected by the suite of past outbreaks.
The stochastic model did not include vaccination of healthcare workers.
We estimated vaccination effectiveness, reported cases, and time from
symptom onset to reporting using studies from West Africa, not DRC. A
strength of our approach was the use of multiple methods to estimate the
outbreak size, even though Gott's Law has not been validated for outbreak
projections.

Our models confirm that the speedup in the conflict area appears to reflect
increased transmission, possibly due to decreased vaccination coverage.
Before October, there was no data to suggest whether the conflict zone would
manifest more transmission and less detection due to inaccessibility of health
services, or less transmission because of reduced mobility, or other outcomes.
The October data suggests that transmission is increased there. It is of course
not clear whether reported counts underrepresent true numbers of cases in these
areas. It may be that a model that explicitly distinguishes transmission rates
in these zones from those in other areas would model the dynamics underlying
these cases more faithfully and produce more accurate projections. Most of the
EVD cases reported in late September and early October occurred in active
conflict zones, and challenges impeding an effective outbreak response have
increased. Strong international partners such as the U.S.\ Centers for Disease
Control and Prevention withdrew their support due to the security concerns.
Although there was rapid deployment of vaccines during this outbreak, we found
that the impact of vaccines on transmission reduction has been limited at best.
Our stochastic model,
which included high, low, and no vaccine coverage scenarios,
was much less likely to fit high coverage scenarios than
those with low or no coverage, especially with more recent data included.
Thus, as
contact tracing efforts faltered and estimates of vaccine coverage became
increasingly unreliable, our stochastic model produced estimates of
transmission rates more consistent with levels of vaccine coverage
lower than target levels of 62\% coverage associated with past
programs of contact tracing and vaccination.


As of October 2018, the current outbreak is ongoing and does not yet appear to
be concluding.
We believe current rates of accumulation of cases will continue at least in the
short term.
We do not see evidence to indicate it will expand to the scale of
the 2013--16 outbreak, although the possibility can not be dismissed.
We believe the increased rate of case detection corresponding to the shift of
transmission into conflict zones is due to increased transmission, probably
driven by reduced ability to detect and vaccinate contacts in those locations.
Even as control efforts falter, case fatality rates have been decreasing during
the outbreak.\cite{RW23} With aggressive supportive care,
experimental therapeutics and high-quality facilities
(\emph{e.g.}\ air-conditioned, individualized),
health-seeking behaviors may reduce transmission potential in
communities that are resistant to control efforts.\cite{Funk177451}




\begin{thebibliography}{10}

\bibitem{RW7}
Butler D.
\newblock Models overestimate Ebola cases.
\newblock Nature. 2014;515(7525):18.

\bibitem{RW2}
Shaman J, Yang W, Kandula S.
\newblock Inference and forecast of the current {West African Ebola} outbreak
  in {Guinea}, {Sierra Leone} and {Liberia}.
\newblock PLoS Curr. 2014;6.

\bibitem{RW3}
Reis J, Yamana T, Kandula S, Shaman J.
\newblock Superensemble forecast of respiratory syncytial virus outbreaks at
  national, regional, and state levels in the United States.
\newblock Epidemics. 2018;S1755-4365(17):30174--3.

\bibitem{RW5}
Yamana T, Kandula S, Shaman J.
\newblock Superensemble forecasts of dengue outbreaks.
\newblock J R Soc Interface. 2016;13(123).

\bibitem{RW17}
Reich N, Lauer S, Sakrejda K, Iamsirithaworn S, Hinjoy S, Suangtho P, et~al.
\newblock Challenges in Real-Time Prediction of Infectious Disease: A Case
  Study of Dengue in {Thailand}.
\newblock PLoS Negl Trop Dis. 2016;10(6):e0004761.

\bibitem{RW1}
Graham M, Suk J, Takahashi S, Metcalf C, Jimenez A, Prikazsky V, et~al.
\newblock Challenges and Opportunities in Disease Forecasting in Outbreak
  Settings: A Case Study of Measles in Lola Prefecture, Guinea.
\newblock Am J Trop Med Hyg. 2018;98(5):1489--1497.

\bibitem{RW11}
Li M, Dushoff J, Bolker B.
\newblock Fitting mechanistic epidemic models to data: A comparison of simple
  Markov chain Monte Carlo approaches.
\newblock Stat Methods Med Res. 2018;27(7):1956--1967.

\bibitem{Funk177451}
Funk S, Camacho A, Kucharski AJ, Lowe R, Eggo RM, Edmunds WJ.
\newblock Assessing the performance of real-time epidemic forecasts: A case
  study of the 2013-16 Ebola epidemic.
\newblock bioRxiv. 2018;Available from:
  \url{https://www.biorxiv.org/content/early/2018/04/24/177451}.

\bibitem{RW23}
{World Health Organization}. Ebola situation reports: Democratic Republic of
  the Congo; 2018.
\newblock Online; accessed 15-October-2018.
\newblock \url{http://www.who.int/ebola/situation-reports/drc-2018/en/}.

\bibitem{Kelly331447}
Kelly JD, Worden L, Wannier R, Hoff NA, Mukadi P, Sinai C, et~al.
\newblock Real-time projections of Ebola outbreak size and duration with and
  without vaccine use in {\'E}quateur, Democratic Republic of Congo, as of May
  27, 2018.
\newblock bioRxiv. 2018;Available from:
  \url{https://www.biorxiv.org/content/early/2018/06/04/331447.1}.

\bibitem{RN711}
Ebola haemorrhagic fever in Sudan, 1976. Report of a WHO/International Study
  Team.
\newblock Bull World Health Organ. 1978;56(2):247--70.
\newblock Available from: \url{https://www.ncbi.nlm.nih.gov/pubmed/307455}.

\bibitem{RN371}
Breman JG, Heymann DL, Lloyd G, McCormick JB, Miatudila M, Murphy FA, et~al.
\newblock Discovery and Description of Ebola Zaire Virus in 1976 and Relevance
  to the West African Epidemic During 2013-2016.
\newblock J Infect Dis. 2016;Available from:
  \url{https://www.ncbi.nlm.nih.gov/pubmed/27357339}.

\bibitem{RN712}
Baron RC, McCormick JB, Zubeir OA.
\newblock Ebola virus disease in southern Sudan: hospital dissemination and
  intrafamilial spread.
\newblock Bull World Health Organ. 1983;61(6):997--1003.
\newblock Available from: \url{https://www.ncbi.nlm.nih.gov/pubmed/6370486}.

\bibitem{RN695}
Georges AJ, Leroy EM, Renaut AA, Benissan CT, Nabias RJ, Ngoc MT, et~al.
\newblock Ebola hemorrhagic fever outbreaks in Gabon, 1994-1997: epidemiologic
  and health control issues.
\newblock J Infect Dis. 1999;179 Suppl 1:S65--75.
\newblock Available from: \url{https://www.ncbi.nlm.nih.gov/pubmed/9988167}.

\bibitem{RN714}
Khan AS, Tshioko FK, Heymann DL, Le~Guenno B, Nabeth P, Kerstiëns B, et~al.
\newblock The reemergence of Ebola hemorrhagic fever, Democratic Republic of
  the Congo, 1995. Commission de Lutte contre les Epidémies à Kikwit.
\newblock J Infect Dis. 1999;179 Suppl 1:S76--86.
\newblock Available from: \url{https://www.ncbi.nlm.nih.gov/pubmed/9988168}.

\bibitem{CDC2001}
{CDC and Ministry of Health: T  Oyok}, Odonga C, Mulwani E, Abur J, Kaducu F,
  Akech M, et~al.
\newblock Outbreak of Ebola Hemorrhagic Fever --- Uganda, August 2000--January
  2001.
\newblock MMWR: Morbidity and Mortality Weekly Report. 2001;50(5):73--77.
\newblock Available from: \url{https://www.cdc.gov/mmwr/PDF/wk/mm5005.pdf}.

\bibitem{RN715}
Outbreak(s) of Ebola haemorrhagic fever, Congo and Gabon, October 2001-July
  2002.
\newblock Wkly Epidemiol Rec. 2003;78(26):223--8.
\newblock Available from: \url{https://www.ncbi.nlm.nih.gov/pubmed/15571171}.

\bibitem{BF2005}
Boumandouki P, Formenty P, Epelboin A, Campbell P, Atsangandoko C, Allarangar
  Y, et~al.
\newblock Prise en charge des malades et des d\'{e}funts lors de
  l'\'{e}pid\'{e}mie de fi\`{e}vre h\'{e}morragique due au virus Ebola
  d'octobre \`{a} d\'{e}cembre 2003.
\newblock Bull Soc Pathol Exot. 2005;98(3):218--223.
\newblock Available from:
  \url{https://www.researchgate.net/publication/280954258_Prise_en_charge_des_malades_et_des_defunts_lors_de_l%27epidemie_de_fievre_hemorragique_due_au_virus_Ebola_d%27octobre_a_decembre_2003.}

\bibitem{WHO2005}
{World Health Organization}.
\newblock Weekly epidemiological record.
\newblock Abonnement annuel. 2005;43(80):369--376.
\newblock Available from: \url{http://www.who.int/wer/2005/wer8043.pdf}.

\bibitem{RN718}
Nkoghe D, Kone ML, Yada A, Leroy E.
\newblock A limited outbreak of Ebola haemorrhagic fever in Etoumbi, Republic
  of Congo, 2005.
\newblock Trans R Soc Trop Med Hyg. 2011;105(8):466--72.
\newblock Available from: \url{https://www.ncbi.nlm.nih.gov/pubmed/21605882}.

\bibitem{RN710}
Rosello A, Mossoko M, Flasche S, Van~Hoek AJ, Mbala P, Camacho A, et~al.
\newblock Ebola virus disease in the Democratic Republic of the Congo,
  1976-2014.
\newblock Elife. 2015;4.
\newblock Available from: \url{https://www.ncbi.nlm.nih.gov/pubmed/26525597}.

\bibitem{RN716}
MacNeil A, Farnon EC, Morgan OW, Gould P, Boehmer TK, Blaney DD, et~al.
\newblock Filovirus outbreak detection and surveillance: lessons from
  Bundibugyo.
\newblock J Infect Dis. 2011;204 Suppl 3:S761--7.
\newblock Available from: \url{https://www.ncbi.nlm.nih.gov/pubmed/21987748}.

\bibitem{WHO2012}
Uganda: Ebola Situation Report.
\newblock Bull World Health Organ. 2013;Available from:
  \url{https://reliefweb.int/sites/reliefweb.int/files/resources/Uganda-Ebola-27august2012.pdf}.

\bibitem{CDC2017}
{Centers for Disease Control and Prevention}.
\newblock Number of Cases and Deaths in Guinea, Liberia, and Sierra Leone
  during the 2014-2016 West Africa Ebola Outbreak. 2017;Online; accessed
  10-May-2018.

\bibitem{RN727}
Wallinga J, Teunis P.
\newblock Different epidemic curves for severe acute respiratory syndrome
  reveal similar impacts of control measures.
\newblock Am J Epidemiol. 2004;160(6):509--16.
\newblock Available from: \url{https://www.ncbi.nlm.nih.gov/pubmed/15353409}.

\bibitem{RN705}
Blumberg S, Lloyd-Smith JO.
\newblock Comparing methods for estimating $R_0$ from the size distribution of
  subcritical transmission chains.
\newblock Epidemics. 2013;5(3):131--45.
\newblock Available from: \url{https://www.ncbi.nlm.nih.gov/pubmed/24021520}.

\bibitem{RN706}
Lloyd-Smith JO, Schreiber SJ, Kopp PE, Getz WM.
\newblock Superspreading and the effect of individual variation on disease
  emergence.
\newblock Nature. 2005;438(7066):355--9.
\newblock Available from: \url{https://www.ncbi.nlm.nih.gov/pubmed/16292310}.

\bibitem{RW13}
Dalziel B, Lau M, Tiffany A, et~al.
\newblock Unreported cases in the 2014--2016 {Ebola} epidemic: Spatiotemporal
  variation, and implications for estimating transmission.
\newblock PLoS Negl Trop Dis. 2018;12(e0006161).

\bibitem{RN734}
Henao-Restrepo AM, Camacho A, Longini IM, Watson CH, Edmunds WJ, Egger M,
  et~al.
\newblock Efficacy and effectiveness of an rVSV-vectored vaccine in preventing
  Ebola virus disease: final results from the Guinea ring vaccination,
  open-label, cluster-randomised trial (Ebola \c{C}a Suffit!).
\newblock Lancet. 2017;389(10068):505--518.
\newblock Available from: \url{https://www.ncbi.nlm.nih.gov/pubmed/28017403}.

\bibitem{RN733}
Dalziel BD, Lau MSY, Tiffany A, McClelland A, Zelner J, Bliss JR, et~al.
\newblock Unreported cases in the 2014-2016 Ebola epidemic: Spatiotemporal
  variation, and implications for estimating transmission.
\newblock PLoS Negl Trop Dis. 2018;12(1):e0006161.
\newblock Available from: \url{https://www.ncbi.nlm.nih.gov/pubmed/29357363}.

\bibitem{gott1993implications}
Gott~III JR.
\newblock Implications of the Copernican principle for our future prospects.
\newblock Nature. 1993;363:315--319.
\end{thebibliography}

\section*{Supplemental Information}

\subsection*{Results}

\subsubsection*{Data Sources}

\begin{table}[h!]
\centering
\small
\begin{knitrout}
\definecolor{shadecolor}{rgb}{0.969, 0.969, 0.969}\color{fgcolor}
\begin{tabular}{p{1.3in}p{1in}p{0.7in}p{0.5in}p{0.9in}p{0.9in}p{0.9in}}
\textbf{Time Period} & \textbf{Country} & \textbf{Reported Count} & \textbf{Time Series Count} & \textbf{Regression?} & \textbf{Stochastic?} & \textbf{Auto-Regression?}\\
Aug--Sep 1976 & DRC* & 318 & 262 & Yes & No & Yes\\
Jun--Nov 1976 & Sudan & 284 & 284 & Yes & No & Yes\\
Aug--Sep 1979 & Sudan & 34 & 34 & Yes & Yes & Yes\\
Dec 1994--Feb 1995 & Gabon & 52 & 49 & Yes & No & Yes\\
May--Jul 1995 & DRC & 315 & 317 & Yes & Yes & Yes\\
Jan--Apr 1996 & Gabon & 37 & 29 & Yes & Yes & Yes\\
Jul 1996--Mar 1997 & Gabon & 60 & -- & No & No & No\\
Oct 2000--Jan 2001 & Uganda & 425 & 436 & Yes & No & Yes\\
Oct 2001--Jul 2002 & Gabon, Republic of the Congo & 124 & 124 & Yes & Yes & Yes\\
Dec 2002--Mar 2003 & Republic of the Congo & 143 & -- & No & No & No\\
Nov--Dec 2003 & Republic of the Congo & 35 & 35 & Yes & Yes & Yes\\
Apr--Jun 2004 & Sudan & 17 & 17 & Yes & Yes & Yes\\
Apr--May 2005 & DRC & 12 & 12 & Yes & Yes & Yes\\
Aug--Nov 2007 & DRC & 264 & 264 & Yes & Yes & Yes\\
Dec 2007--Jan 2008 & Uganda & 131 & 127 & Yes & Yes & Yes\\
Dec 2008--Feb 2009 & DRC & 32 & 32 & Yes & Yes & Yes\\
Jun--Aug 2012 & Uganda & 24 & 24 & Yes & Yes & Yes\\
Jun--Nov 2012 & DRC & 52 & 52 & Yes & Yes & Yes\\
Aug--Nov 2014 & DRC & 66 & 62 & Yes & Yes & Yes\\
Jul--Oct 2014 & Nigeria (offshoot of West African outbreak) & 20 & -- & No & No & No\\
Jan 2014--Jun 2016 & Guinea, Liberia, Sierra Leone & 28,616 & 21,422 & Yes & No & Yes\\
Apr--Jun 2018 & DRC & 53 & 53 & Yes & Yes & Yes\\
\end{tabular}

\end{knitrout}
\caption[]{\label{table:past-outbreaks}
\textbf{Table of past outbreaks by year and country.}
Official reported case counts for each epidemic are given,
including suspected cases (``Reported Count''). 
Case counts for the time series data included in the models
include only probable and confirmed cases (``Time Series Count'').
Case counts for historic outbreaks were pulled from publicly available
literature.%
\cite{RN711,RN371,RN712,RN695,RN714,CDC2001,RN715,BF2005,WHO2005,RN718,RN710,RN716,WHO2012,CDC2017}
Lastly, each historic outbreak's inclusion in the regression,
stochastic, and auto-regression models is enumerated.
\\
*Democratic Republic of Congo (formerly Zaire)
}
\end{table}

Table~\ref{table:past-outbreaks} summarizes the past outbreaks
used as data to inform our models.

\begin{figure}
\centering
\begin{tabular}{ll}
Data as of 8-20-2018 & Data as of 8-27-2018 \\
\includegraphics[width=0.4\textwidth]{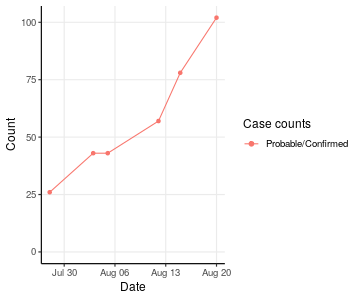} &
\includegraphics[width=0.4\textwidth]{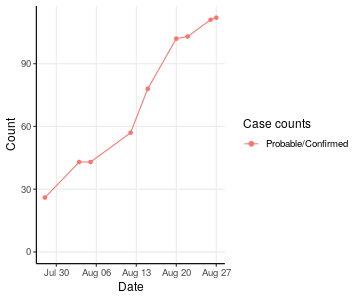} \\
Data as of 9-5-2018 & Data as of 9-15-2018 \\
\includegraphics[width=0.4\textwidth]{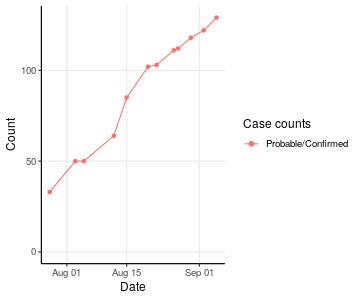} &
\includegraphics[width=0.4\textwidth]{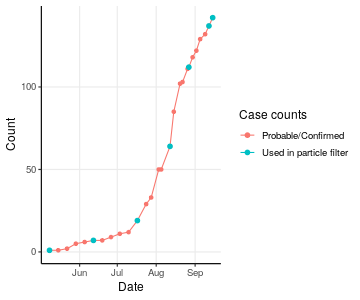} \\
Data as of 10-7-2018 & Data as of 10-13-2018 \\
\includegraphics[width=0.4\textwidth]{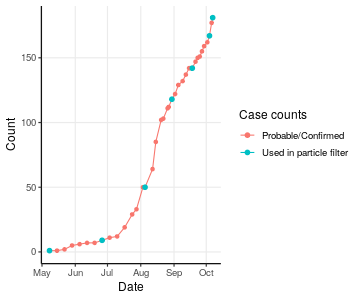} &
\includegraphics[width=0.4\textwidth]{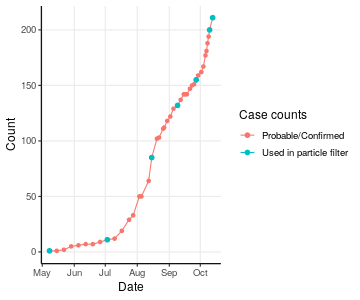}
\end{tabular}
\caption{\label{fig:case-counts}
\textbf{Reported case counts in current outbreak}
by date, in multiple snapshots of available data.
Where not otherwise noted, all case counts shown
were used in the stochastic model's particle filtering step.
}
\end{figure}

We retained snapshots of the set of available case counts
at multiple time points, for use in scoring of
retrospective model projections against known subsequent counts
(Figure~\ref{fig:case-counts}).
In later data sets, due to the larger number of data points,
a subset of the case counts was selected for use in the
stochastic model's particle filtering step, as noted in the figure.

\subsubsection*{Stochastic Model}

\begin{figure}
\centering
\includegraphics[height=0.4\textwidth]{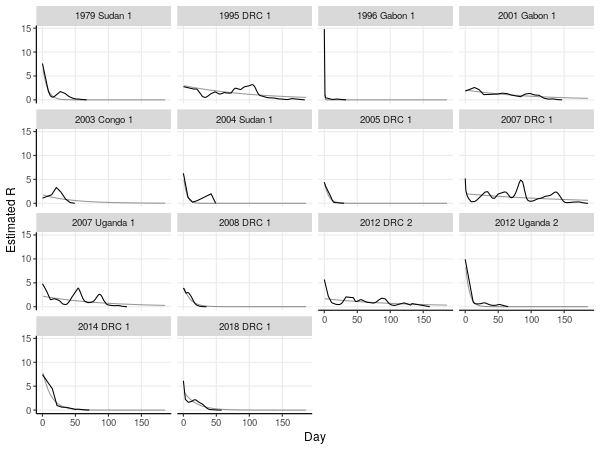}
\caption{\label{fig:R-tau-fits}
\textbf{Estimates of reproduction number $R$}
by day in past Ebola outbreaks. Thin curves are exponentially
quenched curves $R=R_\text{initial} e^{-\tau d}$ fit to each series of $R$ estimates.
}
\end{figure}

\begin{figure}
\centering
\includegraphics[height=0.4\textwidth]{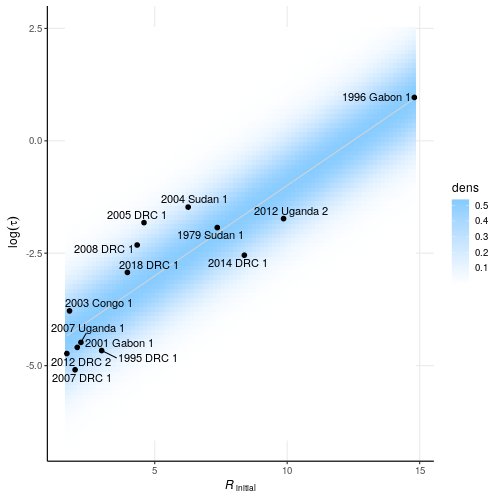}
\caption{\label{fig:R-tau-prior}
\textbf{Distribution of transmission rates sampled for simulation}.
Black dots are pairs $R_\text{initial}$ and quenching rate $\tau$
estimated from past Ebola outbreaks,
and blue cloud is the continuous distribution from which pairs
are sampled for simulation.
}
\end{figure}

Epidemic curves reported for past Ebola outbreaks were
used to estimate time series of effective reproduction number
($R$) by day, which were then fit to an exponential quenching curve
(Figure~\ref{fig:R-tau-fits}).
The parameters $R_\text{initial}$ and $\tau$ estimated by that
curve fitting on past epidemics were then used to create a 
distribution from which values were sampled to parametrize
the stochastic simulation
(Figure~\ref{fig:R-tau-prior}).

\begin{figure}
\centering
\begin{tabular}{ll}
Data as of 8-20-2018 & Data as of 8-27-2018 \\
\includegraphics[width=0.4\textwidth]{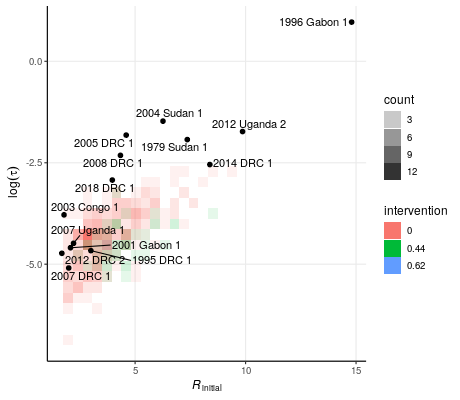} &
\includegraphics[width=0.4\textwidth]{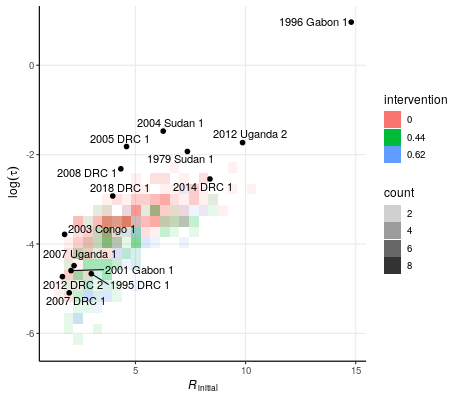} \\
Data as of 9-5-2018 & Data as of 9-15-2018 \\
\includegraphics[width=0.4\textwidth]{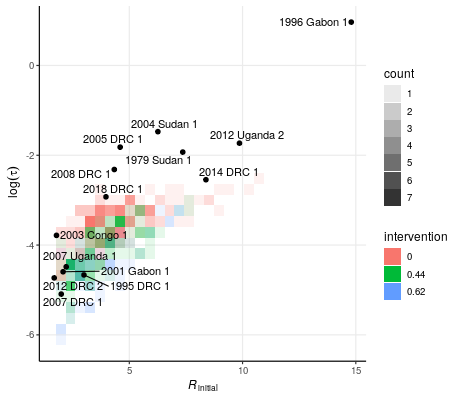} &
\includegraphics[width=0.4\textwidth]{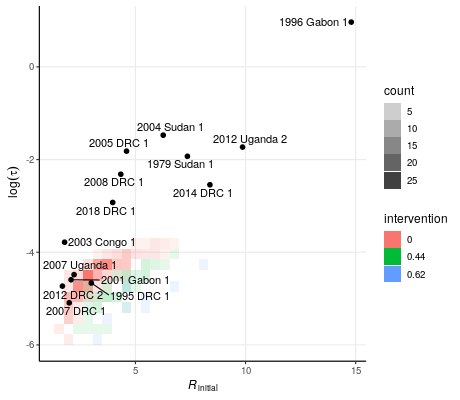} \\
Data as of 10-7-2018 & Data as of 10-13-2018 \\
\includegraphics[width=0.4\textwidth]{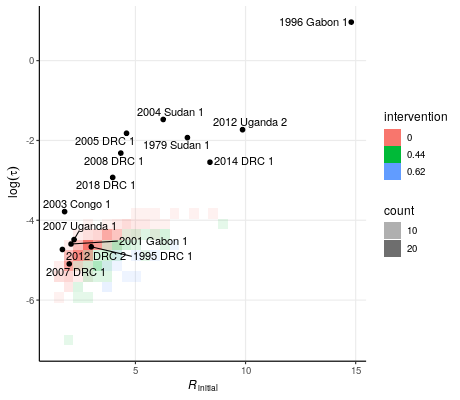} &
\includegraphics[width=0.4\textwidth]{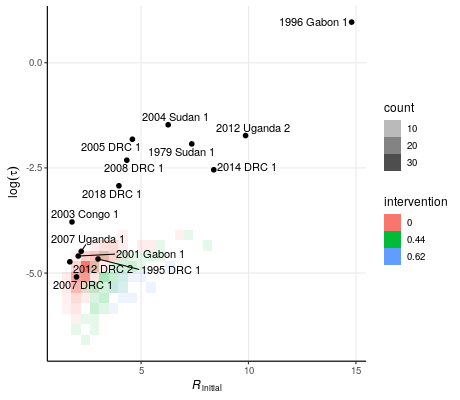}
\end{tabular}
\caption{\label{fig:R-tau-posterior}
\textbf{Transmission rates selected by the particle filtering process},
by vaccine coverage scenario, for successive snapshots of available
case count data.
As in previous figure, black dots for $R_\text{initial}$, $\tau$
pairs estimated for past outbreaks (for comparison),
and colors illustrate the density of $R_\text{initial}$, $\tau$
pairs selected by filtering simulated outbreaks,
classified by level of vaccine coverage.
}
\end{figure}

The $R_\text{initial}$ and $\tau$ parameters driving simulated outbreaks
that were successful in passing the particle filtering step
tended to cluster in particular locations within the assumed distribution
(Figure~\ref{fig:R-tau-posterior}).
In some cases, distinct ranges of $R_\text{initial}$ and/or $\tau$
were selected in conjunction with the different vaccine coverage scenarios. 

For simulation based on cases as of August 20, 320 outbreaks
were retained from 10,196,928 simulated outbreaks
after filtering based on approximate agreement with reported
case counts from the current outbreak.
For the August 27 data set, 320 were retained from 11,622,528;
for September 5, 321 were retained from 6,492,672;
for September 15, 320 from 39,537,792;
and for October 7, 320 from 48,845,376.

\begin{figure}
\centering
\begin{tabular}{ll}
Data as of 8-20-2018 & Data as of 8-27-2018 \\
\includegraphics[width=0.4\textwidth]{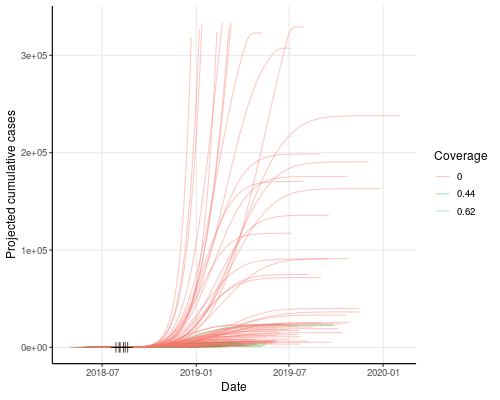} &
\includegraphics[width=0.4\textwidth]{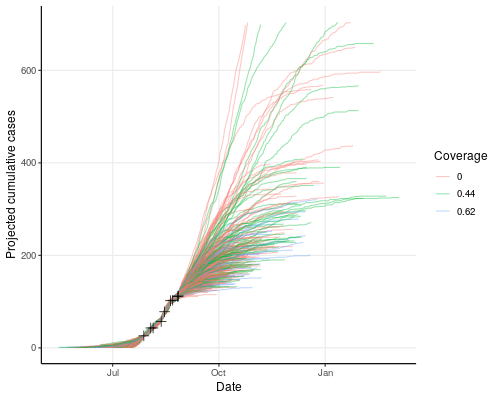} \\
Data as of 9-5-2018 & Data as of 9-15-2018 \\
\includegraphics[width=0.4\textwidth]{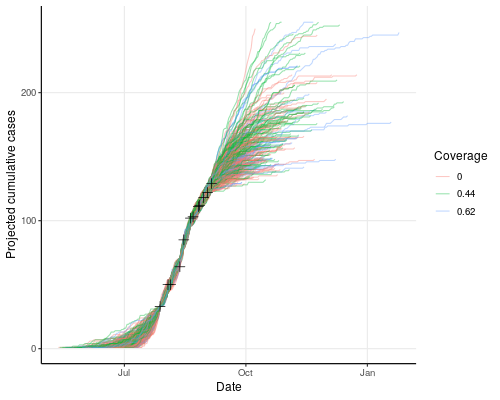} &
\includegraphics[width=0.4\textwidth]{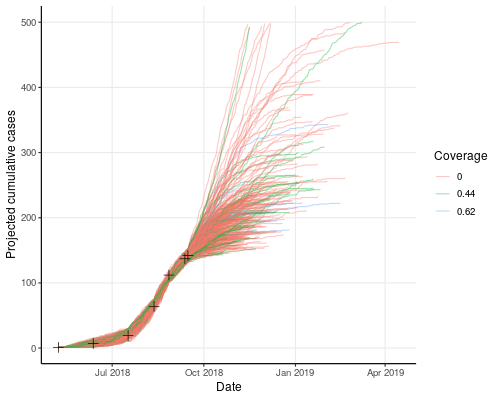} \\
Data as of 10-7-2018 & Data as of 10-13-2018 \\
\includegraphics[width=0.4\textwidth]{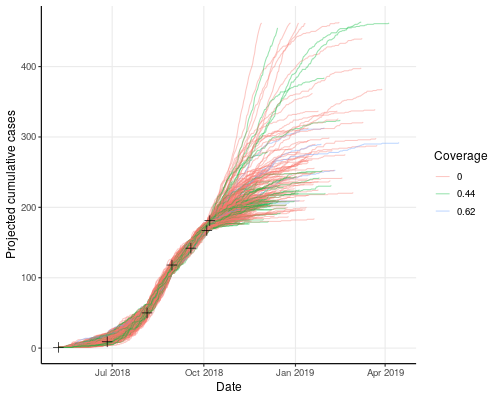} &
\includegraphics[width=0.4\textwidth]{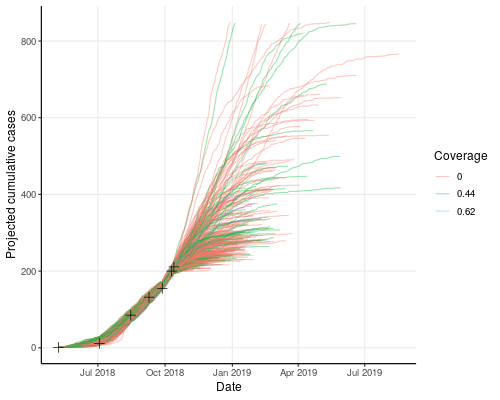}
\end{tabular}
\caption{\label{fig:ribbons}
\textbf{Cumulative case counts by date}
projected by individual realizations of the stochastic model,
by vaccine coverage scenario, using successive snapshots of available
case count data.
The vertical axis is cut off at the upper limit of the 95\% prediction
interval for outbreak sizes, for readability.
}
\end{figure}

The simulations passing the particle filtering step, representing
a distribution of parameter values and vaccine scenarios,
were continued beyond the particle filtering points to
generate a spreading set of projections of case counts at later
dates (Figure~\ref{fig:ribbons}),
which was smoothed to create probabilistic projections
of future case counts at the desired dates.

\FloatBarrier

\subsubsection*{Projections}

\begin{figure}
\centering
\begin{tabular}{ll}
Data as of 8-20-2018 & Data as of 8-27-2018 \\
\includegraphics[width=0.4\textwidth]{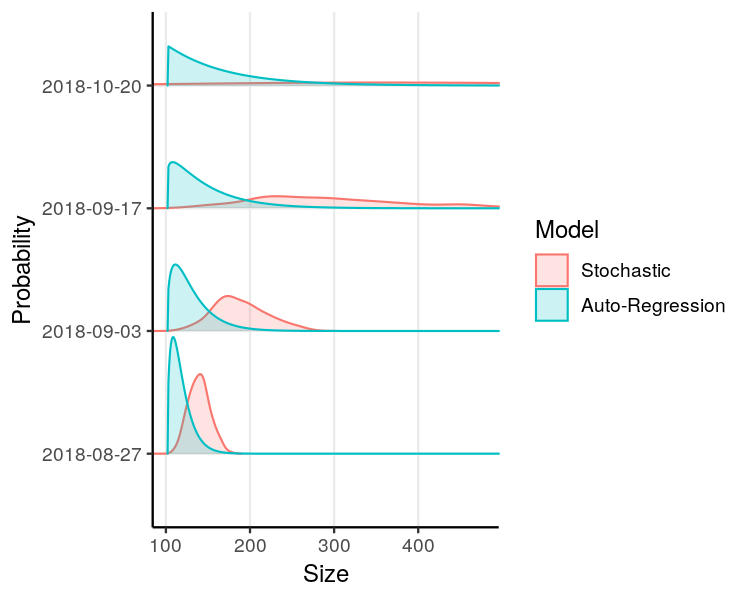} &
\includegraphics[width=0.4\textwidth]{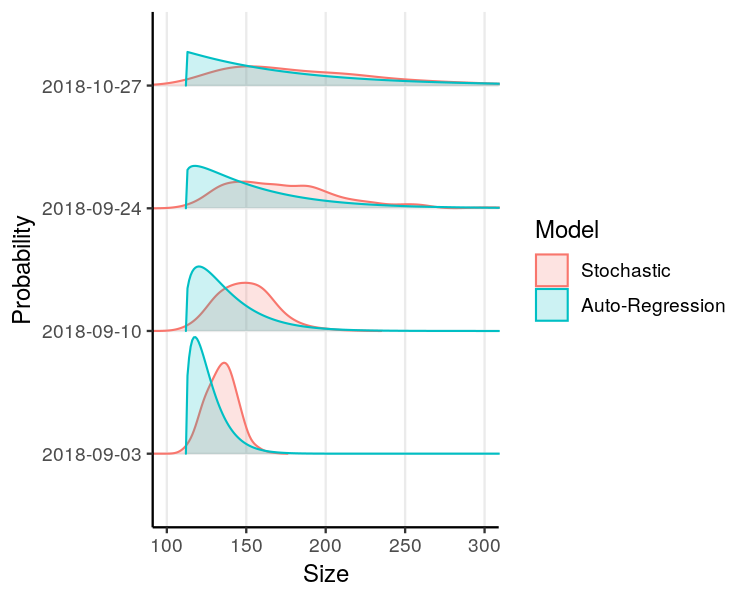} \\
Data as of 9-5-2018 & Data as of 9-15-2018 \\
\includegraphics[width=0.4\textwidth]{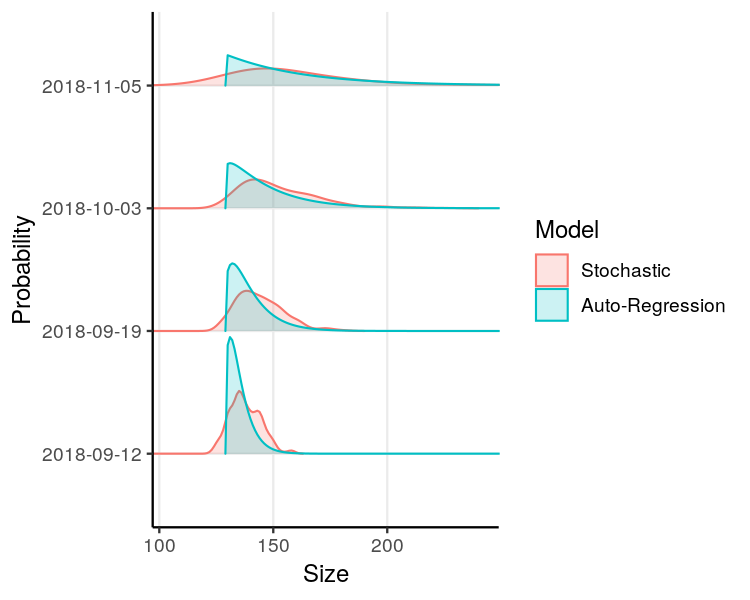} &
\includegraphics[width=0.4\textwidth]{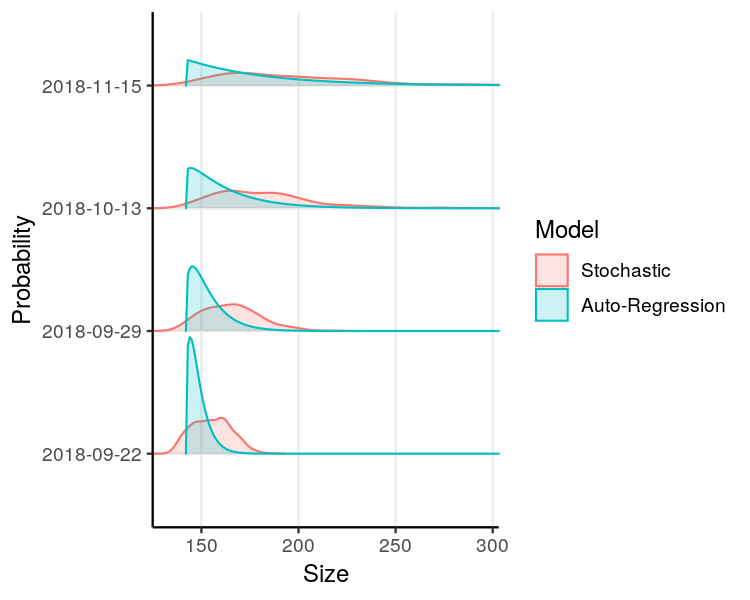} \\
Data as of 10-7-2018 & \\
\includegraphics[width=0.4\textwidth]{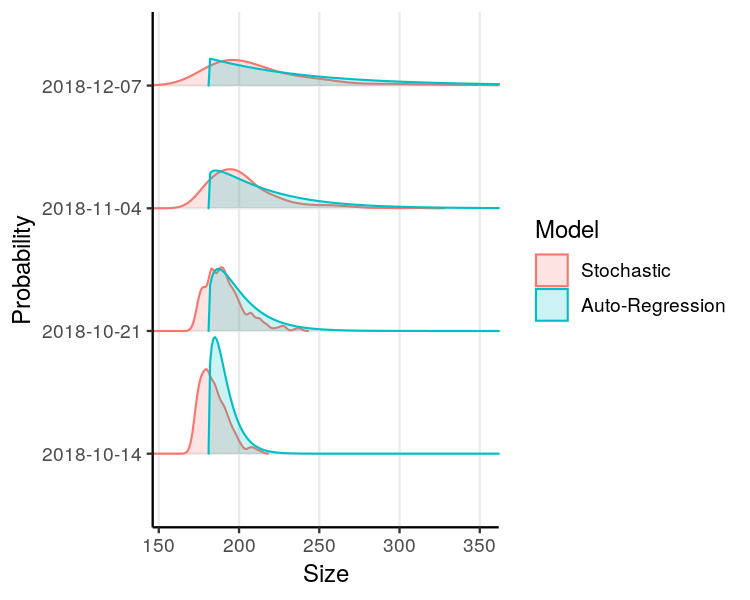} &
\end{tabular}
\caption{\label{fig:short-term-retro-forecasts}
\textbf{Short-term projections} based on past data sets.
}
\end{figure}

\begin{figure}
\centering
\begin{tabular}{ll}
Data as of 8-20-2018 & Data as of 8-27-2018 \\
\includegraphics[width=0.4\textwidth]{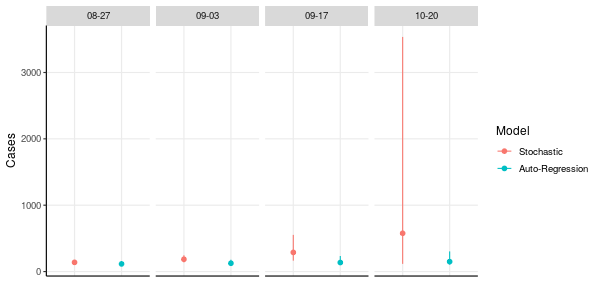} &
\includegraphics[width=0.4\textwidth]{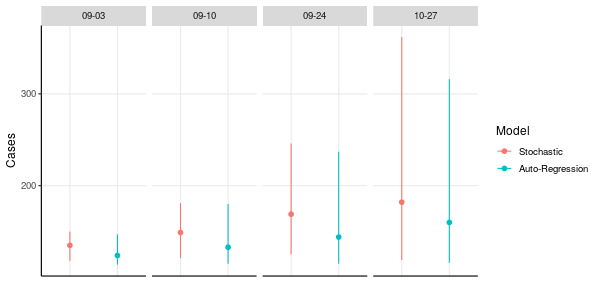} \\
Data as of 9-5-2018 & Data as of 9-15-2018 \\
\includegraphics[width=0.4\textwidth]{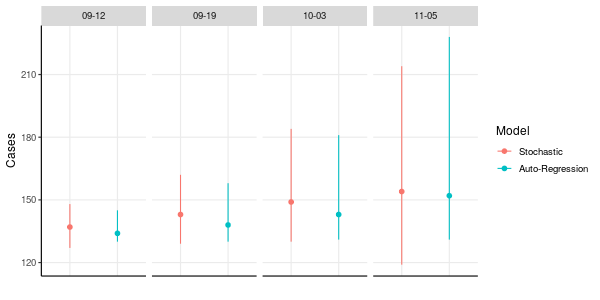} &
\includegraphics[width=0.4\textwidth]{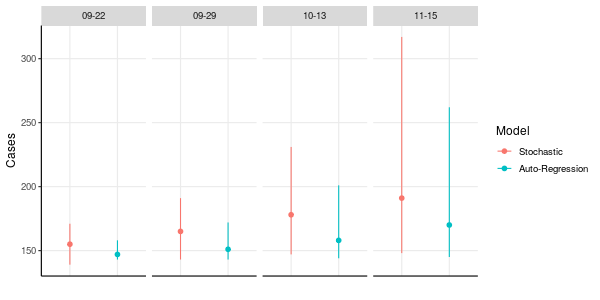} \\
Data as of 10-7-2018 & \\
\includegraphics[width=0.4\textwidth]{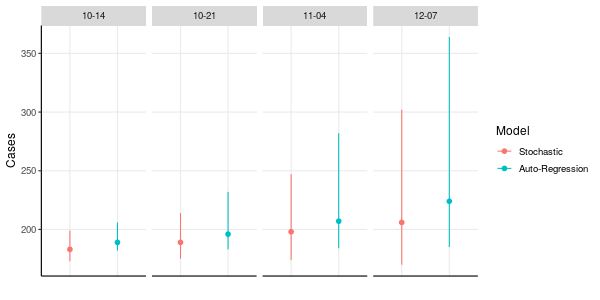} &
\end{tabular}
\caption{\label{fig:short-term-retro-boxplots}
\textbf{Medians and prediction intervals from short-term projections}
based on past data sets.
}
\end{figure}

\begin{figure}
\centering
\begin{tabular}{ll}
Data as of 8-20-2018 & Data as of 8-27-2018 \\
\includegraphics[width=0.4\textwidth]{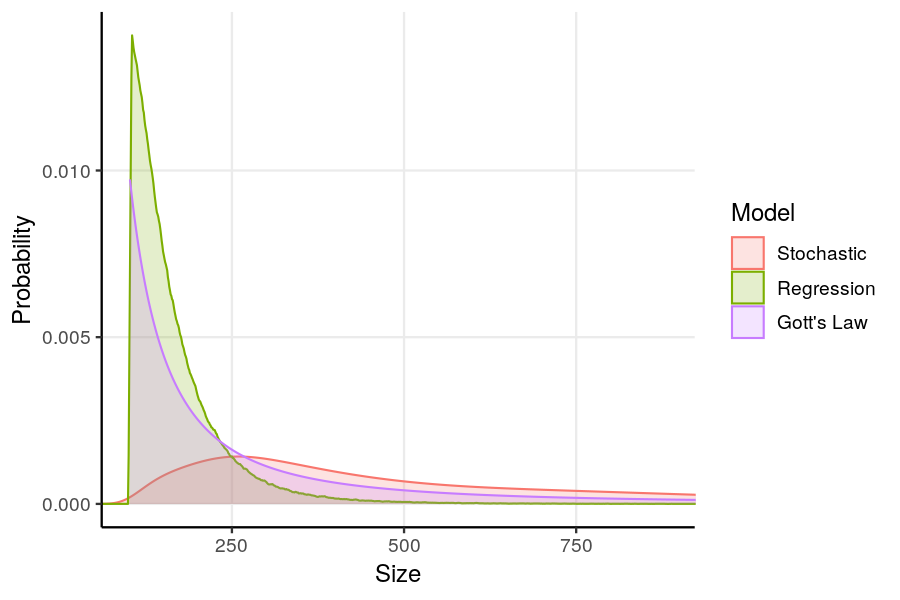} &
\includegraphics[width=0.4\textwidth]{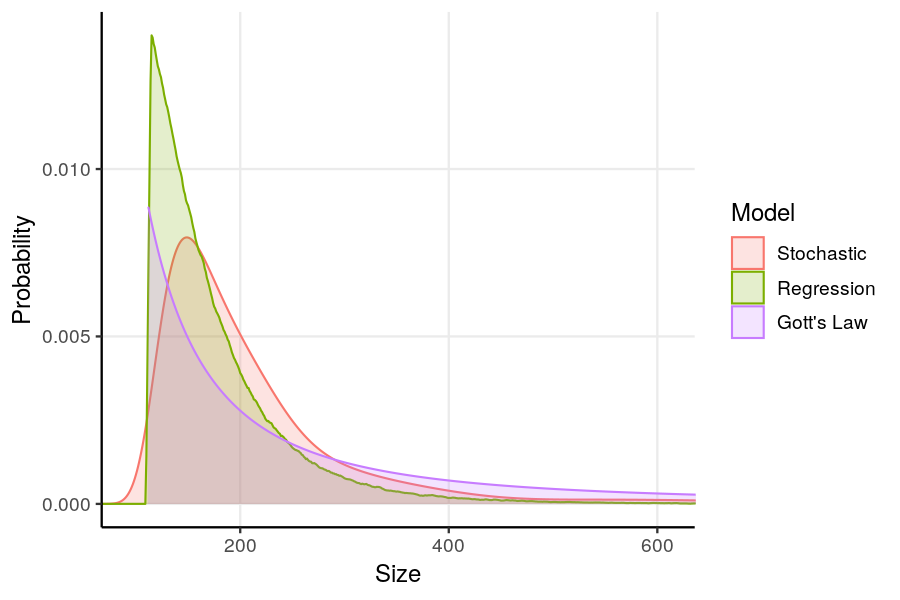} \\
Data as of 9-5-2018 & Data as of 9-15-2018 \\
\includegraphics[width=0.4\textwidth]{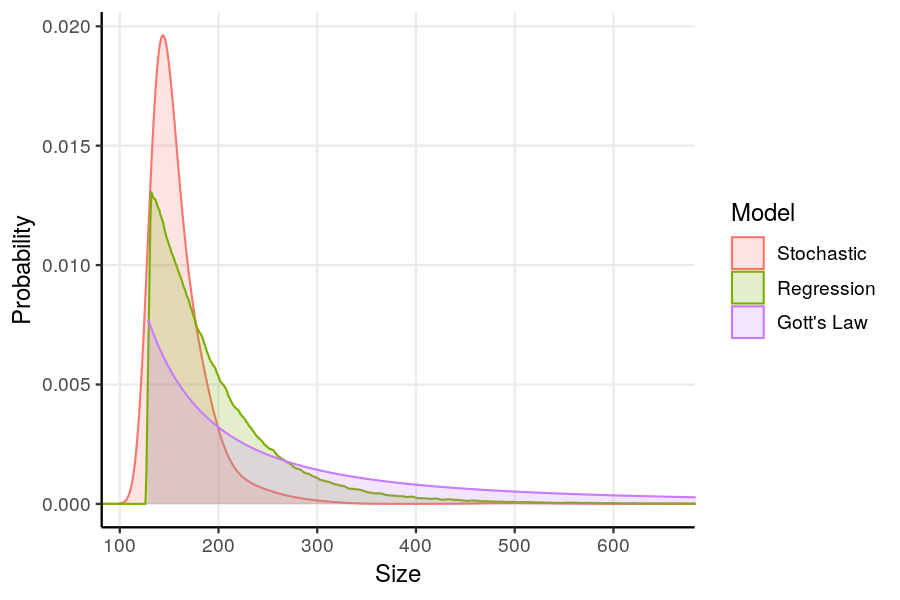} &
\includegraphics[width=0.4\textwidth]{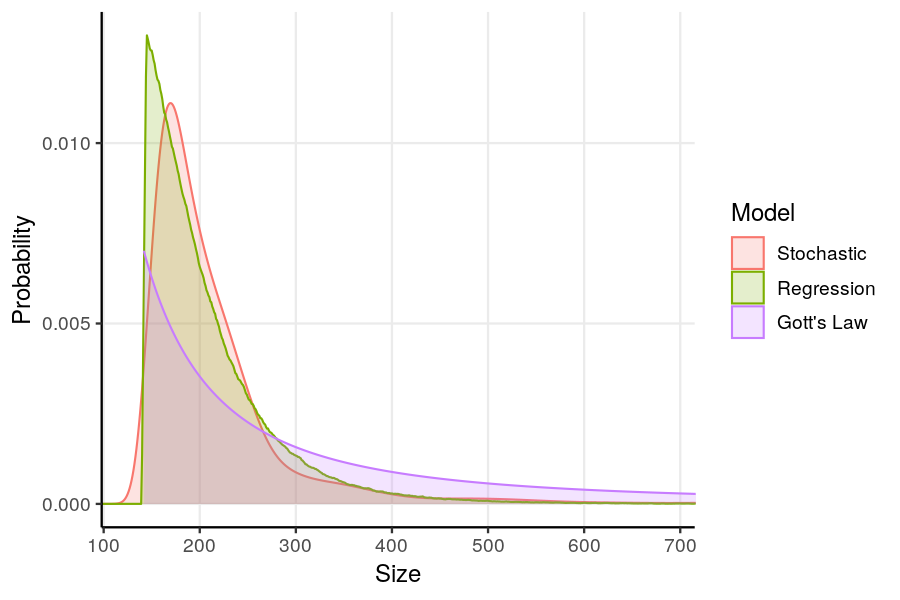} \\
Data as of 10-7-2018 & \\
\includegraphics[width=0.4\textwidth]{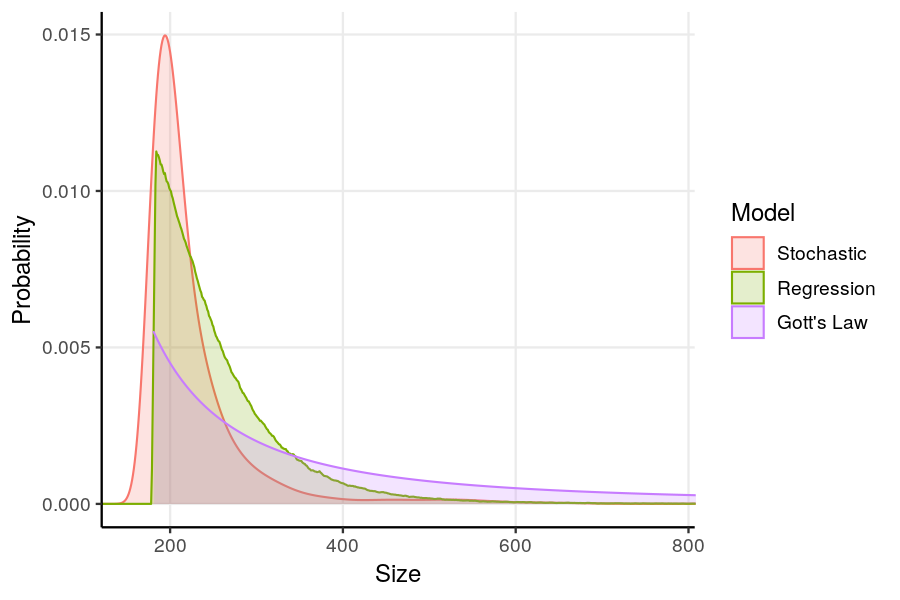} &
\end{tabular}
\caption{\label{fig:final-retro-forecasts}
\textbf{Final outbreak size projections} based on past data sets.
}
\end{figure}

\begin{figure}
\centering
\begin{tabular}{ll}
Data as of 8-20-2018 & Data as of 8-27-2018 \\
\includegraphics[width=0.4\textwidth]{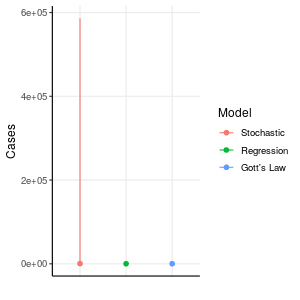} &
\includegraphics[width=0.4\textwidth]{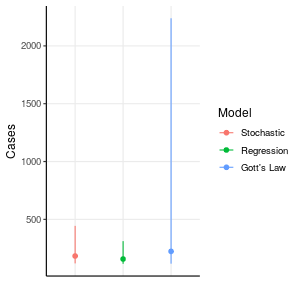} \\
Data as of 9-5-2018 & Data as of 9-15-2018 \\
\includegraphics[width=0.4\textwidth]{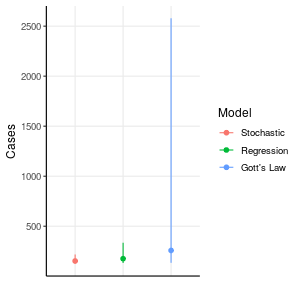} &
\includegraphics[width=0.4\textwidth]{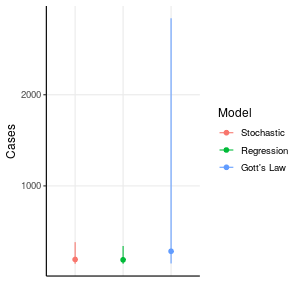} \\
Data as of 10-7-2018 & \\
\includegraphics[width=0.4\textwidth]{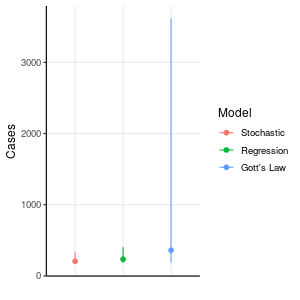} &
\end{tabular}
\caption{\label{fig:final-retro-boxplots}
\textbf{Medians and prediction intervals from final outbreak size projections}
based on past data sets.
}
\end{figure}

We have recorded the projections generated by our models from older
data sets to assess the development of the projections as the
outbreak has progressed,
in Figures~\ref{fig:short-term-retro-forecasts},
\ref{fig:short-term-retro-boxplots},
\ref{fig:final-retro-forecasts},
and~\ref{fig:final-retro-boxplots}.

\FloatBarrier

Table~\ref{table:forecast-stats} summarizes the medians and
95\% prediction intervals produced by each model on the most
recent data set included, and their probabilities of outcomes
exceeding the 2013--2016 outbreak.

\begin{table}[h!]
\centering
\small
\begin{tabular}{lllrrrr}
\textbf{Forecast as of} & \textbf{Forecast} & \textbf{Model} & \textbf{Lower} & \textbf{Median} & \textbf{Upper} & \textbf{Over 28,616}\\
2018-10-13 & 2018-10-20 & Auto-Regression & 214 & 227 & 258 & 0.00000\\
2018-10-13 & 2018-10-20 & Stochastic & 200 & 214 & 243 & 0.00000\\
2018-10-13 & 2018-10-27 & Auto-Regression & 215 & 240 & 307 & 0.00000\\
2018-10-13 & 2018-10-27 & Stochastic & 205 & 226 & 268 & 0.00000\\
2018-10-13 & 2018-11-10 & Auto-Regression & 216 & 259 & 395 & 0.00000\\
2018-10-13 & 2018-11-10 & Stochastic & 208 & 245 & 315 & 0.00000\\
2018-10-13 & 2018-12-13 & Auto-Regression & 217 & 279 & 501 & 0.00000\\
2018-10-13 & 2018-12-13 & Stochastic & 204 & 268 & 428 & 0.00000\\
2018-10-13 & final & Gott's Law & 222 & 421 & 4219 & 0.00526\\
2018-10-13 & final & Regression & 216 & 277 & 485 & 0.00000\\
2018-10-13 & final & Stochastic & 210 & 274 & 632 & 0.00000\\
\end{tabular}
\caption{\label{table:forecast-stats}
\textbf{Table of medians and prediction intervals} of model projections.
}
\end{table}

\end{document}